\AtBeginDocument{%
  \paperwidth=\dimexpr
    1in + \oddsidemargin
    + \textwidth
    + 1in + \oddsidemargin
  \relax
  \paperheight=\dimexpr
    1in + \topmargin
    + \headheight + \headsep
    + \textheight
    + 1in + \topmargin
  \relax
  \usepackage[pass]{geometry}\relax
}

\RequirePackage{fix-cm}
\documentclass[smallextended]{svjour3}       	
\smartqed  
\usepackage{graphicx}

\begin{document}


\title{Authentication Based Solutions to\\ Counterfeiting of Manufactured Goods}
\subtitle{Anti Counterfeiting Solutions for\\ Pharmaceutical Drugs and Currencies.}
\author{Joseph Kilcullen}
\institute{J. Kilcullen \at
		'Moylurg',\\
		Foxford Road,\\
		Ballina,\\
		Co. Mayo,\\
		F26 D9D2,\\
		Ireland\\
              \email{KilcullenJ@gMail.com}
}

\date{Received: date / Accepted: date}

\maketitle

\begin{abstract}
Counterfeiting of manufactured goods is presented as the theft of intellectual property, patents, copyright etc. accompanied by identity theft. The purpose of the identity theft is to facilitate the intellectual property theft. Without it the intellectual property theft would be obvious and the products would be confiscated and destroyed. Authentication solutions, to prevent identity theft, were then developed for the two categories of manufactured goods i.e. goods which can be subjected to destructive screening strategies and goods which cannot e.g. pharmaceutical drugs and currencies, respectively. The solutions developed were found to be analogous to digital signatures. Tamper proof packaging on pharmaceutical drugs is analogous to encryption because it prevents Mallory from interfering with the product. Breaking the tamper proof packaging is a one-way function. Concealed inside the packaging a one-time password, which can be used to authenticate the product over the internet. The name of the authentication website must be common knowledge, just like a public key for authenticating digital signatures. Otherwise the counterfeiters will specify their own authentication website. This solution can be altered for currencies i.e. the one-way function, equivalent to opening the tamper proof packaging, becomes the method of manufacture of the currency.


\end{abstract}

\section{Introduction}
\label{intro}
In a world where technologies proliferate over time, eventually anything you can manufacture, I can manufacture. If true, then over time, there is nothing to stop anything, including paper money, from being counterfeited. As a first thought experiment consider the following: A patented pharmaceutical drug, where the published patent specification is sufficient for anyone to manufacture the drug. Now imagine a generic drug manufacturer utilising the patent specification to manufacture the drug.

Without licencing the patent this manufacturer cannot legally sell the drug. However if they package the drug in identical packaging to the patent holder's product, then, without the rule of law intervening, they can sell their product. The product is such that it actually contains 100\% of the active ingredient. In the same way that a counterfeit handbag is still a handbag, this product is only counterfeit in so far as it is sold with a stolen identity, and stolen intellectual property. The product itself is genuine. The counterfeit handbag is manufactured with a stolen design, but, it too is also a genuine/real handbag. Both of these examples are identity theft counterfeiting i.e. the counterfeit product bypasses trademark law, patent law or copyright law. But the product is not of inferior quality.

Anti-counterfeiting solutions that merely detect defective or substandard products cannot protect against identity theft type counterfeiting. However, identity theft solutions can protect against both substandard products and perfect copy counterfeiting. From this, it follows that we can prevent counterfeiting by preventing manufacturer's identity from being stolen i.e. by preventing identity theft. Identity theft is easily prevented via cryptographic authentication.

This interpretation is supported by earlier work where I found phishing attacks to be the counterfeiting of an identity not the counterfeiting of a website [1].

In the tradition that is academic publication of theories and hypothesis I leave this hypothesis to the academic community to consider. Assuming that this hypothesis is correct I will now proceed to devise cryptographic authentication solutions to protect pharmaceutical drugs and currencies, from counterfeiting. These two products correspond to manufactured goods which can be subjected to destructive testing and goods which cannot. In the context of counterfeiting such strategies are usually referred to as tests, hence 'destructive testing' is a destructive screening strategy.

\section{Authenticating Pharmaceutical Drugs}

Pharmaceutical drugs are packaged in tamper proof packaging which is damaged or destroyed on opening. On receiving a product whose tamper proof packaging has been opened a consumer will either bin the product or return it to the shop. Effectively tamper proof packaging is a game theory signalling strategy. Open packaging does not mean the product has been tampered with, however intact packaging does guarantee that the product has not been interfered with. Consumers understand this so they will not trust the product from opened packaging.

\begin{figure}[!ht]
\centering
\includegraphics[scale=0.3]{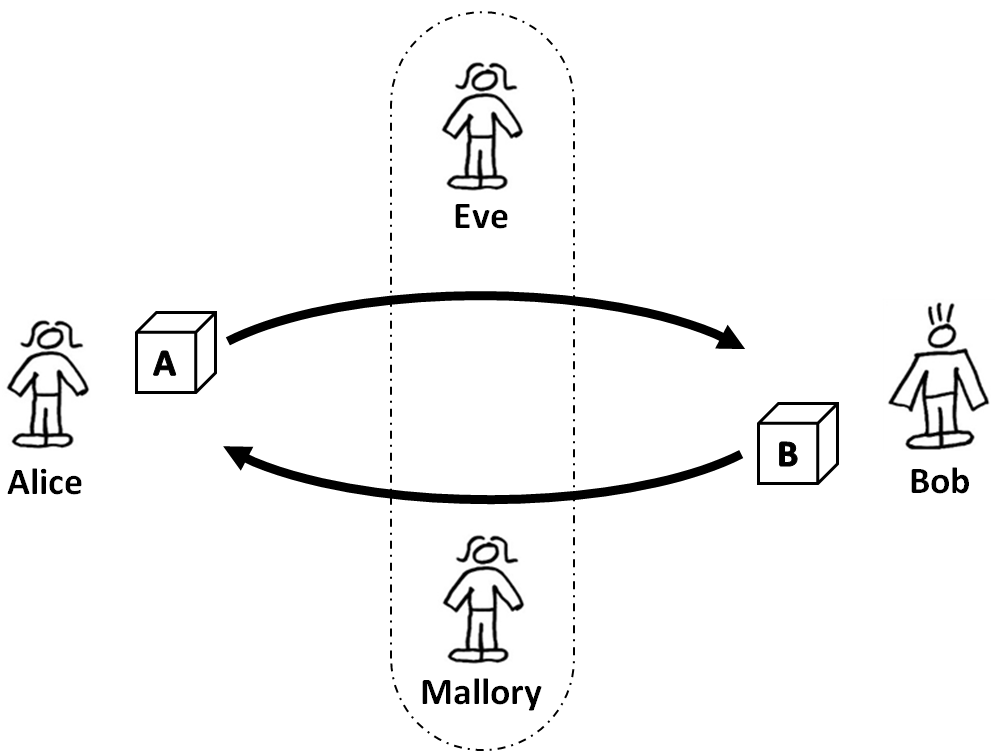}  
\caption{Schematic of classic cryptography actors reinterpreted for counterfeiting of drugs. Eve is shown though she does not play a role in counterfeiting. Tamper proof packaging fulfils the role of encryption because it prevents Eve and Mallory from interfering.}
\label{fig_sim}
\end{figure}

Fig. 1 and Fig. 2 are the typical scenario for cryptography. They are shown here to aid our reinterpretation, of the interaction, as the counterfeiting of pharmaceutical drugs. Box A corresponds to money being exchanged for the pharmaceutical product, box B. Eve is shown though she does not play a role in counterfeiting. Fig. 2, can represent a Phishing attack or the sale of a counterfeit product. In each case Mallory simply 'packages' the fake product in packaging, identical to the original. For Phishing attacks this means creating a website which looks like the original. For counterfeiting of pharmaceutical drugs this means replicating the genuine product's packaging. Tamper proof packaging is analogous to encryption. It prevents Eve and Mallory from interfering. Fig. 2 shows the counterfeiters, Mallory, taking on the role of Bob. This is a consequence of inadequate authentication of Bob.

\begin{figure}[!ht]
\centering
\includegraphics[scale=0.28]{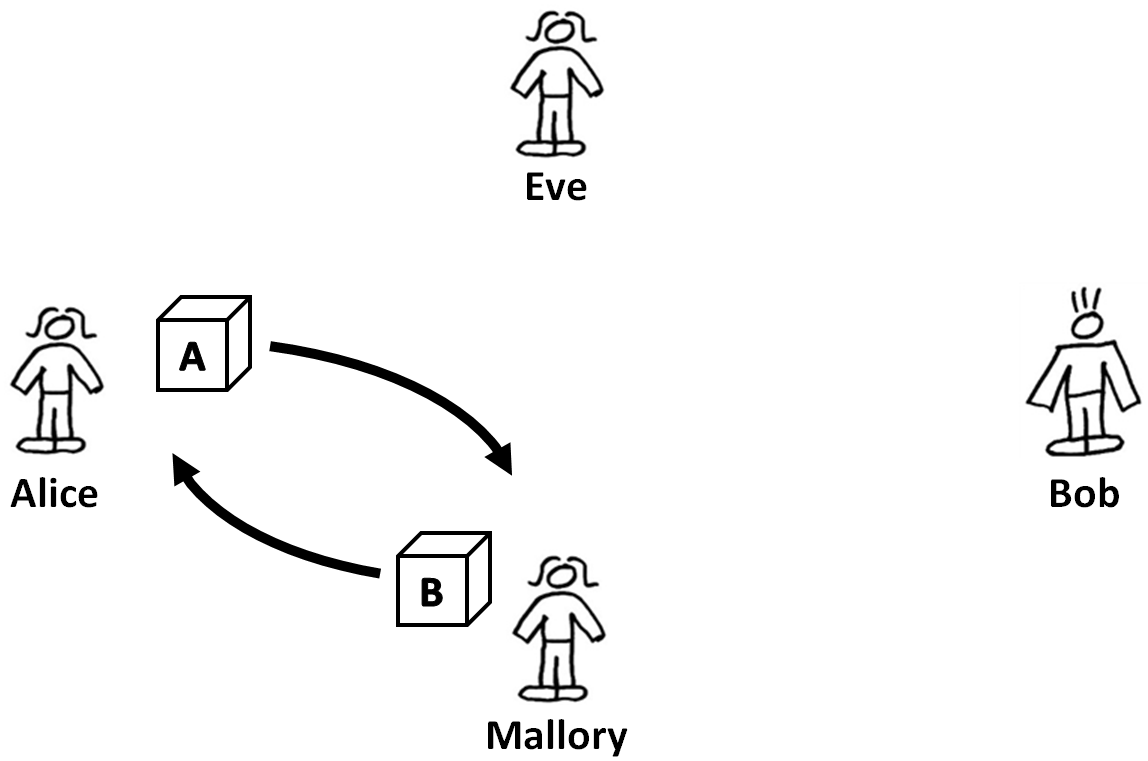}  
\caption{Schematic of Phishing attack scenario, or counterfeit goods scenario. The counterfeiters, Mallory, take on Bob's role. Also they place the product inside tamper proof packaging, ironically, so that Alice knows that nobody has interfered with the product.}
\label{fig_sim}
\end{figure}

TLS in its current form prevents eavesdropping yet it fails to complete Bob's authentication [1]. In the same way tamper proof packaging merely guarantees that the product has got to you safely from the factory. It does not tell whether the package came from Bob or Mallory, from the genuine manufacturer or a counterfeiter. This is analogous to a phishing attack carried out over a TLS connection. We must complete Bob's authentication to ensure it's not counterfeit.

\subsection{Prior Art - Existing Systems \& Solutions}

I do not believe this solution can be patented because of a number of prior art solutions and, at least one, similar patent. A part of this solution which is not prior art (to my knowledge) is documented below in the section titled 'The Common Knowledge Website'. Here follow some examples of prior art:

\vspace{1mm}

\begin{itemize}
  \item Various promotional offers exist where promotional codes (single use authentication passwords) have been concealed inside a product's tamper proof packaging. Kellogg's currently have a Free-Spoon offer where the shared-secret is printed on the inside of the cereal box. Similarly Coca-Cola has had offers where promotional codes (authentication passwords) have been printed on the inside of bottle caps.

  \item Pre-paid mobile phone top-ups. The top-up code is a single use password. Notionally it is concealed until the time of purchase when the shop assistant prints it. (Consider this after studying the solution presented below.)

  \item References [2] and [3] are patent specifications for a very similar solution to the solution outlined below. Their first claim contains an additional clause which is not present in my solution.
\end{itemize}

\vspace{1mm}

A key component of this solution which is not prior art is the 'common knowledge' website, detailed below.

\subsection{Pharmaceutical Drug Authentication}

Here follows an outline of the solution. As stated this is quite similar to references [2] and [3], though I arrived at this solution following the research into phishing attacks [1].

\vspace{1mm}

\begin{itemize}
  \item Tamper proof packaging enclosing the product. This is analogous to encryption because it prevents Mallory from tampering with the product.
  \item Various product identifiers, on the product, such as product name, manufacturer name, batch number, product number etc. 
  \item Collectively the product identifiers should uniquely identify each product. Analogous to a username.
  \item A secret known by the product manufacturer which is concealed inside the tamper proof packaging, with the product. Both the product identifiers and this secret are unique to this product. During manufacturing Bob stores, in a computer database, the identifiers and secret of each product he manufacturers. This secret is equivalent to a password.
  \item A list of instructions, a procedure for the customer to authenticate the product. By, the customer, after opening the tamper proof packaging, communicating to the manufacturer, via website, text message or other, the identifiers and the concealed secret, whereupon:
  \item The manufacturer will compare the customer's product identifiers and product secret to actual identifiers and secrets used by the manufacturer.
  \item Incorrect identifiers and secret indicate a counterfeit. Correct values may indicate a genuine product.
  \item Correct Identifiers and secret: The manufacturer will authenticate the product as genuine if the product has not previously been authenticated. Otherwise the manufacturer should reply with details of the time and date of its original authentication i.e. it's a single use password. If this identifier-secret pair has been used a number of times the manufacturer can communicate this to Alice i.e. a large number of counterfeits with this identifier-secret pair have been manufactured.
  \item If customs, or law enforcement, open a product and test its authenticity they should put the product into a new tamper proof package with new identifiers and a new secret. These should be authenticated through the same website. This is to prevent counterfeiters from using this as a trick. Alternatively pharmaceutical drug companies could include a second pair of identifiers and authentication codes specifically for customs officers, and law enforcement.
\end{itemize}

\vspace{1mm}

Next I will discuss a major flaw with this solution i.e. counterfeiters will place their own website address on their counterfeit.

\subsection{The 'Common Knowledge' Website}

You authenticate a digital signature by (a) using a public key to carry out a one-way function. And (b) you compare the output of that function with what you expected i.e. you 'authenticate the output'.

Reference [4] actually compares a one-way function to breaking a plate\footnote{Humpty Dumpty?}. Opening the tamper proof packaging destroys the packaging i.e. it's a one-way function. 

Like digital signatures, we authenticate the pharmaceutical product by (a) carrying out a one-way function by breaking the tamper proof packaging. And (b) we compare the concealed secret with the manufacturer's secret via a website, or other.

With digital signatures the public key is common knowledge i.e. readily available from a reliable source. In the same way the domain name of the authenticating website should be common knowledge.

That is, the entire pharmaceutical industry should use one website address. Then they publicise that one address. Part of publicising that address includes stating that any product which specifies a different website, for authentication, is a counterfeit. If Mallory, counterfeiters, continue to specify their own authentication website then their products will be clearly visible as counterfeit. Basically Mallory, a counterfeiter, will be forced into using the correct website address. When this happens Alice ends up seeking authentication for counterfeit goods from Bob's website. That authentication process will fail and Alice will know that the product is fake.

\section{Authenticating Currencies}

Three principle changes will allow us to implement the same solution for currencies. First, currencies will be unreproducible physical objects (UPO), as detailed below. Specifically the one-way function, equivalent to opening the tamper proof packaging, becomes the method of manufacture of the currency. Secondly, we need screening strategies to confirm the authenticity of the currency's UPO. And thirdly, there is no need for the common knowledge website because the central bank's name is already common knowledge.

\subsection{Unreproducible Physical Objects}
Consider the objects shown in Fig. 3. Unreproducible physical objects (UPO) are such that nobody can create two UPOs that are identical.

Consider a manufacturing production line. Sheets of glass are laminated and cut into small square tiles. Finally each glass tile is struck with a hammer, shattering the glass. No two tiles will have the same pattern of cracks in the glass. Hence anyone can manufacture UPOs but nobody can counterfeit a specific UPO. Not even the original manufacturer. They are the random output of one-way functions. Unreproducible = Uncounterfeitable.

\begin{figure}[!ht]
\centering
\includegraphics[scale=0.28]{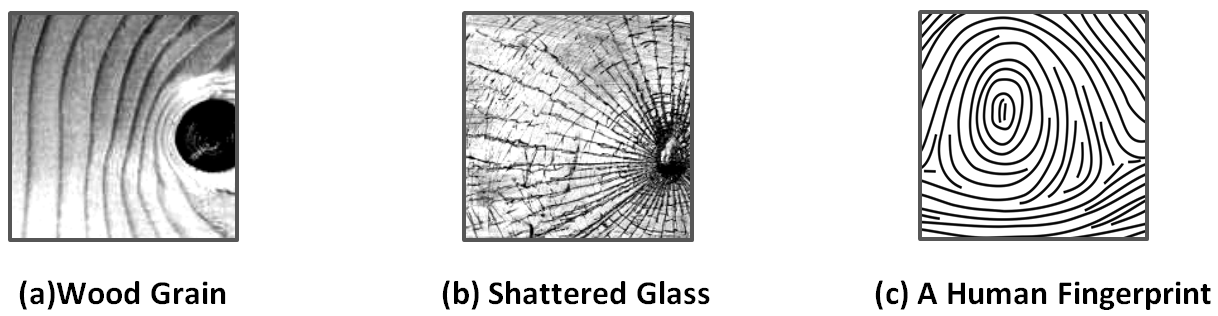}
\caption{Shown are (a) a piece of wood with a grain pattern, (b) a piece of shattered glass and (c) a human fingerprint.}
\label{fig_sim}
\end{figure}

Each UPO created is actually unique. I can try to create two identical shattered tiles but the pattern of cracks will be unique for each tile. This is the whole point of these UPOs. I can manufacture them, but even I cannot counterfeit an existing UPO. In the same way I can grow a tree but the grain pattern, of any given piece of wood, will be unique. Hence (a) in Fig. 3. However there is a problem.

Counterfeiters don't have to 100\% reproduce the product, they can approximate it. In reference [1] I found that approximation of solutions broke more of my anti-counterfeiting solutions than 100\% reproduction. The wood grain pattern could be scanned into a computer, then printed onto a sticker and attached to a rigid wood like base.

In theory we have made progress towards resolving counterfeiting. In practice the counterfeiters will 'up their game' and we find ourselves looking for screening strategies again i.e. we need uncounterfeitable wood, or other.\footnote{Arguable a 'counterfeit human' is more commonly known as a 'human clone'. A natural experiment in human cloning is identical twins. A quick search on the internet indicates that identical twins possess the same DNA, yet have different fingerprints. See Fig. 3 (c).}

If the counterfeiters ability to approximate a UPO can be resolved then attaching a UPO to a product is equivalent to giving that product a unique fingerprint. A unique identity. Or, in the case of currencies, the currency will be a UPO itself.

\subsection{Authenticating Currency UPOs}

Non-destructive screening strategies are still needed to confirm the authenticity of the currency. However their purpose is to confirm that the fingerprint is genuine. If shattered glass is used, how do we know a printout of the break pattern hasn't been placed on a sheet of glass of the correct weight? Some fingerprint scanners can be fooled with basic printouts of the correct fingerprint.\footnote{Google 'mythbusters fingerprint hack'.}

Effectively our non-destructive screening strategies are directed at preventing the UPO fingerprint from being counterfeited. We authenticate UPO currencies by (a) subjecting the currency to non-destructive screening strategies to ensure the fingerprint has not been counterfeited. And (b) we read the fingerprint and serial numbers. Finally (c) we utilise a website/other to seek confirmation that this fingerprint-serial number combination is authentic.

\subsection{Coins Example}
Coins could be made out of glass. Flaws like tiny bubbles would be difficult to position during manufacture. As such imperfections would make each coin unique and impossible to replicate. Whatever process is used to draw a picture and a coin value, could be used to add a unique serial number. Authentication would involve authenticating the physical properties like refractive index and examining the coin's unique physical flaws i.e. fingerprint. Finally the central bank, for that currency, can authenticate that 'serial number - fingerprint' combination.

\subsection{Paper Money Example}
Here follows an outline of one possible UPO solution for paper money. The thought experiments which led to this solution have been omitted.

\vspace{1mm}

\begin{itemize}
  \item Some, or all, of the bank note should be a UPO i.e. some part of the note should be possible to manufacture but impossible to precisely duplicate. Like the shattered glass example.
  \item The UPO portion of the bank note should be amenable to one, or more, non-destructive tests which will authenticate it. For example, does a screening strategy exist to ensure that an approximation has not been printed onto the note, to look like the genuine UPO? That is, a good choice of UPO cannot be printed without a test existing to 'reveal' that it is a counterfeit, an approximation.
  \item When measured the physical characteristics should generate a unique fingerprint, a unique 'id' of this unique bank note.
  \item Digital signatures are not practical. It's more straight forward to print a unique serial number onto each note and authenticate over the internet.\footnote{The discussion on printing digital signatures onto money has been omitted. Consider A3 sized banknotes to facilitate printing a digital signature onto the note?}
  \item Metal incorporated into the bank note, similar to the metal strip, could be used provided the magnetic properties could still be measured/tested.
\end{itemize}

\vspace{1mm}

\par All that remains is to identify an appropriate UPO. The following is suggested as an example of what that UPO might look like. Other mechanisms could also be developed.

\vspace{1mm}

\begin{itemize}
  \item In the middle of paper money, like the metal strip, sprinkle a large number of tiny pieces of metal over a specific area e.g. 1cm\textsuperscript{2}. Or incorporate metal into the tiny pieces of cotton that are used to create the paper money, cotton money i.e. sprinkle tiny metal flecks over the entire bank note.

  \item Different metals with different magnetic properties should be used. Precious metals like gold or silver could be used but they would only be there for their unique magnetic properties. Not as a commodity.

  \item The random scattering of a random selection of metals should generate a unique magnetic map. By unique I mean an extremely low probability that two notes would have the same magnetic fingerprint.

  \item The idea is that magnetic properties should vary over the surface area. According to the type and arrangement of metal flecks present.

  \item Any material or combination of materials which would help make the fingerprint impossible to counterfeit could be used. Including metal flecks with, difficult to discern, combinations of metals. Materials like Graphene may also have a distinct effect on magnetic properties.

  \item The objective is to have several variables which will vary over the surface of the UPO. They should be easy to measure and test to ensure they have not been counterfeited. The precise combination should generate a unique fingerprint. The density of metal flecks would be determined by the ability to measure the magnetic response and capacity of the paper to physically hold that many metal flecks in place. The area covered, the types of metals used, the density of flecks and spread/location of those metal flecks should, collectively, interfere with a magnetic field in a unique manner i.e. a fingerprint. Some machine/device should be developed which would interrogate the magnetic properties over the surface area of the fingerprint.

  \item Some rock candy sweets are manufactured by making one massive sweet with the text, or picture, constructed from large lumps of candy.\footnote{Search on YouTube for 'How its made - hard candy'.} Then it is rolled into long bars. As it is rolled the text, or picture, shrinks. In the same way long bars of metal could be combined to form one large bar with the same arrangement of metals as a specific UPO. At that point a wafer thin slice could be taken off and placed into a counterfeit bank note. If the density of metal flecks in the genuine fingerprint allows gaps. Then shining a bright light through the paper will reveal these gaps. A scanned image, taken like this, could be incorporated into the fingerprint. In this way the metal bar process similar to manufacturing candy will not work. Incorporating metal flecks into entire bank notes will also solve this problem.

  \item Tests other than measurement of magnetic properties could be used e.g. what if magnetic ink was used to print a replica fingerprint. Shining light through such notes may appear different to the original. Hence our fingerprint should be devised to be resilient to such counterfeiting attempts. A backlit scan could be incorporated into the fingerprint i.e. the fingerprint, itself, can be the output of tests to authenticate the physical fingerprint.

\end{itemize}

\par Authenticating the bank note consists of:

\vspace{1mm}

\begin{itemize}

  \item First authenticate the UPO. By definition we should have chosen a UPO which is amenable to having its authenticity tested. As described above.

  \item Once we're satisfied that the UPO is genuine we measure, read, the fingerprint off of it. It's possible that we already have the fingerprint from the authentication of the UPO.

  \item Finally we utilise a website to communicate with the central bank that issued the note. We transmit the serial number on the bank note and the fingerprint measured from the note's UPO. The website should then respond, indicating the authenticity of any note with that combination of serial number and fingerprint.

\end{itemize}


Some remaining points:
\begin{itemize}
  \item The concept of the common knowledge website is not relevant here. It's obvious who issued the bank note. Also there is the possibility to limit access to the authenticating website to banks and law enforcement, via password.
  \item Even without authentication a UPO should help protect against counterfeiting.
  \item Internet authentication could be restricted to large denomination bank notes.
\end{itemize}

\par Effectively we are in an arms race. This solution simply forces the counterfeiters to counterfeit the UPOs. While our definition of UPOs works we are still in an arms race. The candy example is given to demonstrate how ingenious tactics will be employed to reproduce our UPO. The arms race has not gone away. The tactics and ground rules have changed. Since every financial transaction cannot have bank notes scanned and authenticated. Law enforcement will still be involved in identifying counterfeits and tracking them back to their origins.

\section{Conclusion}
Regular cryptography requires both encryption and decryption, as do digital signatures. When working with manufactured goods one-way functions can be used, like breaking tamper proof packaging. The fact that these acts cannot be undone can be used to our advantage. The combination of tamper proof packaging, a concealed secret and a common knowledge website, for authentication, is analogous to a digital signature. Our ability to communicate with the legitimate manufacturer allows us to communicate the secret, from within the product packaging, to seek confirmation that the product is genuine, analogous to utilising a public key to authenticate a digital signature.

Where products cannot be damaged e.g. currencies, we must make the product itself the output of a random one-way function. Anyone can manufacture currencies of this kind, but only the original central bank will authenticate its notes as genuine. This process is equivalent to the pharmaceutical drug example except the currency is manufactured by the random output of a one-way function. Provided the physical fingerprint cannot be counterfeited it can be authenticated by the legitimate manufacturer, thus preventing counterfeiting.


\end{document}